\documentstyle[12pt]{article}  
\newcommand{\be}{\begin{equation}}
\newcommand{\ee}{\end{equation}}
\newcommand{\ba}{\begin{eqnarray}}
\newcommand{\ea}{\end{eqnarray}}

\begin{document}
\begin{center}
{\bf\Large{2D-gravity and the Hamilton-Jacobi formalism}}
\end{center}
\begin{center} {\bf Dumitru Baleanu}\footnote[1]{ On leave of absence from 
Institute of Space Sciences, P.O BOX, MG-23, R 76900
Magurele-Bucharest, Romania,
E-mail: dumitru@cankaya.edu.tr}
and
{\bf Yurdahan G\"uler}\footnote[2]{E-Mail:~~yurdahan@cankaya.edu.tr}
\end{center}
\begin{center}
Department of Mathematics and Computer Sciences, Faculty of Arts
and Sciences, Cankaya University-06530, Ankara , Turkey
\end{center}
\begin{abstract}
Hamilton-Jacobi  formalism is used to study 2D-gravity and its 
SL(2, R) hidden 
symmetry. If the contribution of the surface term is considered the 
obtained results coincide with those given by the Dirac and 
Faddeev-Jackiw approaches.
\end{abstract}

\newpage

 \section {Introduction}

  Two dimensional gravity models were subjected to an intense
 investigation during last years due to its relation with
 string theory and its potential applications to a better understanding
 of classical and quantum properties of gravity models 
\cite{barbashov, polyakov,ooguri, katanaev, witten, fillipov,jackiw, 
strobl2000, kummerreport}.
 Polyakov \cite{polyakov} proved that an induced 2D-gravity has a 
hidden $SL(2,R)$ symmetry. This symmetry was analysed in the 
context of the canonical and extended Hamiltonian formalisms 
\cite{manvelian90,2dabdala, barcelos94, devecchi94} as well as in the 
context of improper gauge transformations \cite{subir94}.
Despite of a huge number of publications on the above topics still a final
 conclusion and consensus on the main problems is missing.  
 
 Recently, Hamilton-Jacobi (HJ) formalism was investigated in the 
context of  strings and p-Branes  \cite{hjstringsbrane}, strongly
coupled gravitational systems \cite{hjsalopek},nonholonomic-constrained 
systems with second-class
constraints and Proca's model \cite{hong}   
and the quantum  HJ formulation was obtained from the 
equivalence principle \cite{hjquantum}.

 A new method of quantization of the system with constraints based on the
$Carath{\acute e}odory$'s equivalent Lagrangians method \cite{car} was 
initiated in \cite{g5,g6} and  developed in \cite{p11, p12,  
gb1,gb2, gb3}.   
The connection between the  integrability  conditions 
and Dirac's consistency conditions \cite{dirac} were established 
\cite{p13}. 
 The importance of the surface terms \cite{mark} was analyzed and 
the 
relation  between Batalin-Fradkin-Tyutin \cite{fad} and HJ  
formalism \cite{gb2} was traced.     
 The  advantage  of HJ formalism
is that we have no difference between first and second class
constraints and we do not need gauge-fixing term because
the gauge variables are separated in the processes of constructing  an
integrable system of total differential equations.
 The action of the formalism is suitable
for the  path integral quantization method of the 
constrained systems.
However the role of the surface term for HJ formalism requires a 
deep analysis especially when we are dealing with gravity.

  For these reasons the application of HJ formalism to 2D-gravity and the 
investigation of its SL(2,R) hidden symmetry is an interesting issue.

  The plan of the paper is as follows:

In sec. 2 HJ formalism is presented.
 2D-gravity is analyzed using HJ formulation in sec.3
 In sec. 4 conclusions are given.

\section{Hamilton-Jacobi formalism}
Let us assume that the Lagrangian L is singular and the Hessian 
matrix has the rank n-r. The "Hamiltonians" to start with are

\be\label{doi} H_{\alpha}^{'}=H_{\alpha}(t_{\beta},q_{a},p_{a})
 +p_{\alpha}, \ee where $\alpha,\beta=n-r +1,\cdots,n$,$a=1,\cdots
n-r$. The usual Hamiltonian $H_0$ is defined as

\be\label{unu} H_{0}=-L(t,q_{i},{\dot q_{\nu}},{\dot q_{a}=w_{a}})
+p_{a}w_{a} + {\dot
q_{\mu}}p_{\mu}\mid_{p_{\nu}=-H_{\nu}},\nu=0,n-r+1,\cdots,n. \ee
which is independent of $\dot q_{\mu}$. Here $\dot
q_{a}={dq_{a}\over d\tau}$, where $\tau$ is a parameter.
 The equations of motion are obtained as total differential equations
in many variables as follows

\ba\label{(pq)}
&dq_{a}&= {\delta H_{\alpha}^{'}\over\delta
p_{a}}dt_{\alpha},
dp_{a}=-{\delta H_{\alpha}^{'}\over\delta
q_{a}}dt_{\alpha},\cr
&dp_{\mu}&=-{\delta H_{\alpha}^{'}\over\delta
t_{\mu}}dt_{\alpha}, \mu=1,\cdots, r , \ea

\be\label{(z)} dz=(-H_{\alpha} + p_{a}{\delta
H_{\alpha}^{'}\over\delta p_{a}})dt_{\alpha}, \ee where
$z=S(t_{\alpha}, q_{a})$ and ${\delta H_{\alpha}^{'}\over\delta x}$
represents the variation of $ H_{\alpha}^{'}$ with respect to x.
The variations of constraints will produce a complete set of 
"Hamiltonians"
and some of them are not in the form given in (\ref{doi}) or they 
are divergence as in the
Proca's case. In order to maintain the physical significance of multi
HJ formulation we are forced to modify the "Hamiltonians" 
by reducing or
extending the initial phase space.
We can adapt the above formalism to 2D-gravity considering the components 
of 
the metric as fields and taking into account that the canonical 
Hamiltonian is zero due to the reparametrization invariance. 

\section{Hamilton- Jacobi formulation of 2D-gravity}
The action proposed by Polyakov for 2D-gravity  is \cite{polyakov}
\be\label{poly}
S=-{1\over 2}
\int\sqrt{-g}[
g^{\alpha\beta}\partial_{\alpha}\phi\partial_{\beta}\phi
+\alpha R\phi]d\tau d\sigma.
\ee

After some integrations by parts this action gives expression for the 
Lagrangian density as 
\ba\label{lunu}
&L=&{1\over 2} \sqrt{-g}g^{\mu\nu}{\partial_{\mu}\phi}\partial_{\nu}\phi +
{\alpha\over 
2\sqrt{-g}}(\epsilon^{\mu\alpha}\partial_{\mu}\phi)(\epsilon^{\nu\beta}\partial_{\beta}g_{\mu\nu})\cr
&-&{\alpha\phi\over 8\sqrt{-g}}
g^{\alpha\beta}\epsilon^{\rho\sigma}\epsilon^{\mu\nu}(\partial_{\mu}g_{\alpha\rho})
(\partial_{\nu}g_{\beta\sigma}),
\ea

where the determinant g is given by

\be
g= g_{00}g_{11}-g_{01}^{2}.
\ee
               
Here $g_{00}, g_{01}$ and $g_{11}$ represent the independent fields.

In order to simplify the form of the Hamiltonian we add to (\ref{lunu}) a 
surface term \cite{manvelian90}. The new Lagrangian density has the form

\be\label{ldoi}
{\tilde L}=L + \partial_{\mu}({\phi\over 2\sqrt{-g}g_{11}}\epsilon^{\mu\nu}\epsilon^{\alpha\beta}g_{1\alpha}\partial_{\nu}g_{1\beta})
\ee

 or explicitly

\ba\label{hjtilda}
{\tilde L}&=&{1\over 2\sqrt{-g}}((-g_{11}\dot{\phi}^2 +
2g_{01}\dot\phi\phi^{'}-g_{00}{\phi^{'2}})
+\alpha\left({\dot g_{11}}{\dot\phi} -
2g_{01}^{'}{\dot\phi}+g_{00}^{'}\phi^{'}\right) \cr
&+&{\alpha g_{01}\over g_{11}}(g_{11}^{'}{\dot\phi}-
{\dot g_{11}}\phi^{'})).
\ea
The overdots and primes denote time and space derivatives, respectively.
The above expression leads us to the 
 canonical momenta
corresponding to $\phi$ and $g_{11}$  as  
\be
\pi_{\phi}={(g_{01}{\phi}^{'}-g_{11}\dot\phi)\over\sqrt{-g}}+
{{\alpha}\over 2\sqrt{-g}}({\dot g_{11}}-2g_{01}^{'}+
{g_{01}g_{11}^{'}\over g_{11}}),
\ee
\be
\pi^{11}={\alpha\over 2\sqrt{-g}}(\dot\phi-{g_{01}\over 
g_{11}}{\phi}^{'}).
\ee

The other momenta  $\pi_{00},p_{01}$ are zero.

The canonical Hamiltonian is expressed as 
\be
H_{c}=\int dx(-{\sqrt{-g}\over g_{11}}\Phi_1+{g_{01}\over 
g_{11}}\Phi_{2}),
\ee
where $\Phi_1$ and $\Phi_2$ are

\be\label{fi1}
\Phi_{1}={1\over 2}\left({\phi^{'2}}- 4({g_{11}\pi^{11})^2\over\alpha^2}
-{4\over\alpha}(g_{11}\pi^{11})\pi_{\phi}
-{\alpha g_{11}^{'}\phi^{'}\over g_{11}}+2\alpha\phi^{''}\right),
\ee

\be\label{fi2}
\Phi_{2}=\pi_{\phi}\phi^{'}-2g_{11}\pi^{11'}-\pi^{11}g_{11}^{'}.
\ee

Thus, the "Hamiltonians" 
which are the basic components of the method, are expressed as
\be\label{hcj}
 H_{0}^{'}=p_{0} + H_{c},
H_{1}^{'}=\pi_{00}, H_{2}^{'}=p_{01}.
\ee
Here $g_{00}$ and $g_{01}$ are gauge
variables.
Consistency conditions require the variations of 
$H_{0}^{'},H_{1}^{'},H_{2}^{'}$. These variations lead us to new 
constraints
\be\label{hcj1}
H_{3}^{'}=\Phi_{1},H_{4}^{'}=\Phi_{2}. 
\ee
Further variations of $H_{3}^{'}$ and $H_{4}^{'}$ do not give independent 
constraints.
The above results are in agreement with those obtained by 
Dirac's canonical formalism \cite{manvelian90}. 

\subsection{$SL(2,R)$ symmetry in the light-cone gauge }
To exibit the $SL(2, R)$ symmetry we introduce the light cone variables
$x^{\pm}$ which are given as 
 $x^{\pm}={1\over\sqrt{2}}(x^{0}\pm x^{1})$. 
Besides, choosing the metric tensor as
\\
\\
\\

\ba
g_{\mu\nu}&=&\left(\matrix{
g_{++}& g_{+-} \cr
g_{-+}& g_{--} \cr}\right)\cr
&=&\left(\matrix{
h& -1 \cr
-1& 0 \cr}\right), 
\ea
where h is a field, the Lagrangian density becomes

\be
L=\partial_{+}\phi\partial_{-}\phi +{1\over 
2}h(\partial_{-}\phi)^2 -{\alpha\over 2}\partial_{-}\phi\partial_{-}h,
\ee
where $\partial_{\pm}={1\over\sqrt{2}}(\partial_{0}\pm \partial_{1})$.    
Treating $x^{+}$ as  time, the momenta corresponding to  $\phi$ and h are
defined as
\be
P={\partial_{-}\phi},
\Pi=0.
\ee
This formulation gives  the canonical Hamiltonian density 
$H_{c}= P\partial_{+}\phi +\Pi h - L$ as

\be
H_{c}=\partial_{+}\phi(P-{\partial_{-}\phi})
+\Pi h -{1\over 2}h(\partial_{-}\phi)^{2}
+{\alpha\over 2}\partial_{-}\phi\partial_{-}h
\ee

Thus, the basic "Hamiltonians" 
to start with are
\ba\label{haha}
&H_{0}^{'}=&p_{0}-
h\partial_{-}\phi\partial_{-}\phi
+\alpha\partial_{-}\phi\partial_{-}h,\cr   
&H_{1}^{'}&=P-{\partial_{-}\phi}, H_{2}^{'}=\Pi    
\ea
Again, the consistency conditions give a new constraint $H_{3}^{'}$ as
\be\label{h3}
H_{3}^{'}= (\partial_{-}\phi)^{2}+\alpha\partial_{-}^{2}\phi  
\ee
The next step is to analyze the  variation of $H_{3}^{'}$.
>From (\ref{haha}) we conclude that 
\be\label{pe}
dP=(h{\partial_{-}^{2}\phi}-{\alpha\over 2}\partial_{-}^{2}h)d\tau
\ee
 Imposing $dH_{1}^{'}=0$ and  using (\ref{pe})   we found the following 
constraint 
\be\label{patru}
H_{4}^{'}={\partial_{-}^{3}h}.
\ee 
The "Hamiltonians" satisfy the following algebra:
\ba\label{sisica}
&\{H_{2}^{'}(x^{-}),H_{4}^{'}(x^{-})\}&=\{H_{4}^{'}(x^{-}),H_{2}^{'}(x^{-})\}=
\partial^{3}_{-}\delta(x^{-}-x^{-'}),\cr
&\{H_{1}^{'}(x^{-}),H_{1}^{'}(x^{-})\}&=-2\partial_{-}\delta(x^{-}-x^{-'}),\cr
&\{H_{1}^{'}(x^{-}),H_{3}^{'}(x^{-})\}&= 
(2\partial_{-}^{'}\phi\partial_{-}-
\alpha\partial_{-}^{2})\delta(x^{-}-x^{-'}),\cr
&\{H_{3}^{'}(x^{-}),H_{1}^{'}(x^{-})\}&= 
(2\partial_{-}^{'}\phi\partial_{-}+
\alpha\partial_{-}^{2})\delta(x^{-}-x^{-'}).
\ea

We would like to mention that in \cite{barcelos94} the same constraints   
were obtained by using Faddeev-Jackiw formalism \cite{fadjack}.
 Algebra (\ref{sisica}) describes a system which is not integrable. 
 Hence we would like to transform it to an integrable one.  
Solving (\ref{patru}) we obtain 
\be\label{gaugehh}
h(x^{-},x^{+})=B_{1}(x^{+})+2x^{-}B_{2}(x^{+})+(x^{-})^{2}B^{3}(x^{+}),
\ee
where $B_{1}(x^{+}),B_{2}(x^{+}),B^{3}(x^{+})$ are arbitrary functions of
$x^{+}$.
 
On the other hand since $H_{c}$ and $H_{3}^{'}$ are the "Hamiltonian" 
densities
we observed that they are proportional up to a surface term.
Taking into account this  result
and using (\ref{gaugehh}) we obtain 
\ba\label{hoson}
&H_{0}^{'}&=p_{0}+B_{1}(x^{+})(\partial_{-}\phi)^{2} +B_{3}(x^{+})[
(x^{-}\partial_{-}\phi)^{2}+\alpha\phi-\alpha\partial_{-}\phi x^{-}]\cr
&+&2B_{2}(x^{+}){1\over 
2}[2x^{-}(\partial_{-}\phi)^{2}-\alpha\partial_{-}\phi].
\ea  

which becomes

\ba\label{hamfin}
&H_{0}^{'}&=p_{0}+B_{1}(x^{+})(P\partial_{-}\phi ) + 
+2B_{2}(x^{+}){1\over
2}[2x^{-}\partial_{-}\phi P -\alpha P]\cr
&+& B_{3}(x^{+})[
(x^{-})^{2}\partial_{-}\phi P +\alpha\phi-\alpha P 
x^{-}]
\ea
with the use of the constraints $H_{1}^{'}=0$.
This expression makes it possible to define quantities
\ba\label{leuri}
&L^{1}(x^{+})&=\int P\partial_{-}\phi dx^{-},\cr
&L^{2}(x^{+})&={1\over
2}\int{[2x^{-}P\partial_{-}\phi  -\alpha P]}dx^{-},\cr
&L^{3}(x^{+})&=\int {[(x^{-})^{2}P\partial_{-}\phi  +\alpha\phi-\alpha P 
x^{-}}]dx^{-}. 
\ea
which fulfill the $SL(2,R)$ algebra on the surface of constraint 
$\int{Pdx^{-}}=0$. 

This result coincides with that obtained 
 by using Dirac's canonical formalism, 
in \cite{manvelian90}  

\section{Conclusions}
The surface term plays an important role for HJ formalism of constrained
systems. This role becomes crucial when a Hamiltonian represents a total
divergence.
  In this paper we applied the HJ formalism to  the 2D-gravity model. 
The analysis of 2D- gravity in the light-cone coordinates is quite 
unusual for HJ formalism.
The starting point was a theory with two "Hamiltonians" and two gauge 
variables. Imposing the integrability conditions 
we found the same constraints as obtained by  using Dirac and 
Faddeev-Jackiw formalisms.  
SL(2,R) symmetry arose in the HJ formalism after we obtained  the reduced 
phase 
space and eliminated the surface terms in order to find an integrable 
system of total differentiable equations.

\section {Acknowledgments}
One of the authors (DB) would like to thank R. Jackiw and F. Devecchi
for their interesting comments and suggestions, and  T. Strobl for reading 
the manuscript.
This work is partially supported
by the Scientific and Technical Research Council of Turkey.

\end{document}